\documentclass[aps,prl,twocolumn,showpacs,amsmath,amssymb,floatfix,superscriptaddress,notitlepage]{revtex4-2}
\usepackage{color}
\usepackage{bbm} 
\usepackage{graphicx} 
\usepackage{dcolumn} 
\usepackage[utf8]{inputenc}
\usepackage{graphicx}
\usepackage{epstopdf}
\usepackage{amsthm}
\usepackage{amsmath}
\usepackage{empheq}
\usepackage{bbm} 
\usepackage{amssymb}
\usepackage[usenames,dvipsnames]{xcolor}
\usepackage{cases}
\usepackage{latexsym}
\usepackage[colorlinks=true,citecolor=Cerulean,linkcolor=RubineRed,urlcolor=Cerulean]{hyperref}

\usepackage{cleveref}

\newcommand{\gs}{E_0}

\newcommand{\id}{\mathbbm{1}} %Identity
\newcommand{\tr}[1]{\operatorname{\textnormal{Tr}}\left( {#1} \right)} 
\newcommand{\ket}[1]{\left|#1\right\rangle}
\newcommand{\bra}[1]{\left\langle#1\right|}

\newcommand{\avg}[1]{\left\langle#1\right\rangle}

\newcommand{\tqsl}{\tau_\textnormal{anneal}}

\newcommand{\tqslone}{\tau_{\textnormal{anneal}1}}
\newcommand{\tqsltwo}{\tau_{\textnormal{anneal}2}}
\newcommand{\tqslthree}{\tau_{\textnormal{anneal}3}}

\newcommand{\tf}{{t_f}}

\begin{document}

\title{Lower Bounds on Quantum Annealing Times}

\author{Luis Pedro Garc\'ia-Pintos}
\email{lpgp@lanl.gov}
\affiliation{Joint Center for Quantum Information and Computer Science,  
University of Maryland, College Park, Maryland 20742, USA} 
\affiliation{Joint Quantum Institute, University of Maryland, College Park, Maryland 20742, USA} 
\affiliation{Theoretical Division (T4), Los Alamos National Laboratory, Los Alamos, New Mexico 87545, USA}

\author{Lucas T. Brady}
\affiliation{Quantum Artificial Intelligence Laboratory, NASA Ames Research Center, Moffett Field, California 94035, USA}
\affiliation{KBR, 601 Jefferson St., Houston, TX 77002, USA}

\author{Jacob Bringewatt}
\affiliation{Joint Center for Quantum Information and Computer Science,  
University of Maryland, College Park, Maryland 20742, USA} 
\affiliation{Joint Quantum Institute, University of Maryland, College Park, Maryland 20742, USA}

\author{Yi-Kai Liu}
\affiliation{Joint Center for Quantum Information and Computer Science,  
University of Maryland, College Park, Maryland 20742, USA} 
\affiliation{Applied and Computational Mathematics Division, National Institute of Standards and Technology, Gaithersburg, MD 20899, USA}

\date{\today}

\begin{abstract}
The adiabatic theorem provides sufficient conditions for the time needed to prepare a target ground state. While it is possible to prepare a target state much faster with more general quantum annealing protocols, rigorous results beyond the adiabatic regime are rare. Here, we provide such a result, deriving lower bounds on the time needed to successfully perform quantum annealing. The bounds are asymptotically saturated by three toy models where fast annealing schedules are known: the Roland and Cerf unstructured search model, the Hamming spike problem, and the ferromagnetic $p-$spin model. Our bounds demonstrate that these schedules have optimal scaling. Our results also show that rapid annealing requires coherent superpositions of energy eigenstates, singling out quantum coherence as a computational resource.
\end{abstract}

\maketitle

\textit{Introduction.---} Generic computational tasks can be mapped to finding the ground state of a Hamiltonian. This is the basis for quantum annealing and adiabatic quantum computing~\cite{annealingPRE1998,
Hauke_2020,
AlbashRevModPhys2018, farhi2000quantum}. 
In these approaches, a computational protocol consists of initializing a system in an easy-to-prepare ground state of a Hamiltonian $H_0$. Thereafter, a time-dependent evolution is performed where the Hamiltonian transitions from $H_0$ to a Hamiltonian $H_1$ whose ground state provides the solution to the desired problem. That is, the system is driven by 
\begin{align}
\label{eq:Ham}
H(t) = (1-g_t) H_0 + g_t H_1,
\end{align}
where the `annealing schedule' $g$ satisfies $g_0 = 0$ and $g_\tf = 1$, and  
$\tf$ is the total duration of the process.

If the transition $H_0 \longrightarrow H_1$ is slow enough, the adiabatic theorem~\cite{Born1928} ensures that the final state is close to the ground state $\big\vert\gs^\tf\big\rangle$ of $H(\tf) = H_1$, in which case the protocol performs the desired computation. 
More precisely,  the system remains close to the ground state at all times if  $\tf \nobreak \geq \nobreak  T_\textnormal{adiab}$, for $T_\textnormal{adiab} \nobreak \sim  \nobreak \theta/\Delta^2$, where $\theta = \max_t \left\|  \tfrac{d}{d(t/\tf) } H(t) \right\|$, $\| \cdot \|$ denotes the spectral norm, and $\Delta$ is the minimum energy gap between the instantaneous ground state and first excited state of $H_t$ over the whole schedule (tighter bounds can also be found in Refs.~\cite{jansen2007bounds,AlbashRevModPhys2018}). 
This mechanism is as powerful as standard quantum computation~\cite{Aharonovuniversal2008}. 
Throughout this work, we use $\sim$ to denote leading order terms, up to multiplicative constants.

The condition 
$\tf \geq T_\textnormal{adiab}$ is  a sufficient one to perform adiabatic computation. 
Necessary conditions were also
derived in Refs.~\cite{BoixoSommaPRA2010,ChenArxiv2022}. 
In Ref.~\cite{BoixoSommaPRA2010}, it was shown that an adiabatic annealing process requires at least a time 
$\tau_{\textnormal{adiab}} \nobreak \sim \nobreak L / \Delta  $, where $L \coloneqq \int_0^\tf \left\| \tfrac{d}{dt} \ket{\psi_t} \right\| dt$ is the length of the adiabatic path that the state $\ket{\psi_t}$ of the system takes in state space. An algorithm for achieving such $\sim 1/\Delta$
scaling by making use of an oracle for the gap is given in Ref.~\cite{jarret2018quantum}. Ref.~\cite{ChenArxiv2022} used bounds on the speed of adiabatic evolution to derive necessary conditions $t_f \geq \tau_{\textnormal{adiab}}$ on adiabatic annealing times.
Throughout this Letter, we denote lower bounds on the time of an annealing process by $\tau_*$ (for some descriptive $*$); we use $T_\textnormal{adiab}$ to denote timescales that ensure that the process is adiabatic.

However, adiabaticity is not a requirement for annealing---it is simply a (powerful yet demanding!) condition that guarantees the successful preparation of the desired ground state with bounded error.
This does not exclude the existence of non-adiabatic annealing schedules that take the system to the desired target state more quickly. This is the motivation behind a plethora of popular (and somewhat overlapping) approaches including the quantum approximate optimization algorithm (QAOA)~\cite{farhi2014quantum,Barraza_2022}, diabatic quantum annealing~\cite{crosson2021prospects}, counterdiabatic driving~\cite{berry2009transitionless, demirplak2003adiabatic,odelin2019shortcuts} and optimal control~\cite{odelin2019shortcuts}. Unfortunately, these approaches are often heuristic with limited theoretical guarantees on performance.

In this Letter, we derive saturable lower bounds on the time necessary for the system to approach a desired target state $\big\vert\gs^\tf\big\rangle$ for quantum annealing protocols. In this way, we find general conditions that constrain how fast annealing can be successfully performed, including beyond the adiabatic regime.

\emph{Asymptotically saturable bounds on annealing times.---} 
We consider
\begin{align}
    C_1(\rho_t) \nobreak \coloneqq \nobreak \min_{\sigma_t} \| \rho_t \nobreak-\nobreak \sigma_t \|_1,
\end{align}
as a measure of energy coherence of the system's state $\rho_t \nobreak = \nobreak \ket{\psi_t}\!\bra{\psi_t}$~\cite{CoherencePRL2014,CoherencePRA2016,
CoherenceRevModPhys2017}, 
where $\sigma_t$ is diagonal in the eigenbasis of $H(t) \nobreak = \nobreak\sum_j E_j^t \ket{E_j^t}\!\bra{E_j^t}$ 
and $\| A \|_1 \nobreak \coloneqq \nobreak \tr{ \sqrt{A A^\dag} }$ denotes the trace norm.
It holds that $C_1(\rho_t) \nobreak\leq\nobreak C_{l_1}(\rho_t)$, where $C_{l_1}(\rho_t) \nobreak\coloneqq \nobreak \sum_{j \neq k} \left| \big\langle E_k^t \big\vert \rho_t \big\vert E_j^t \big\rangle \right|$ is another popular measure of coherence~\cite{CoherencePRA2016}.

Without loss of generality, we take the ground state energies of $H_0$ and $H_1$ to be zero, and we denote their time evolving energy expectation values as $\langle H_0 \rangle_t \nobreak \coloneqq \nobreak  \tr{\rho_t H_0 }$ and $\langle H_1 \rangle_t \nobreak \coloneqq \nobreak  \tr{\rho_t H_1}$. 
The aim of a successful annealing schedule is to maximize the probability $p_{0,\tf}$ to end in the ground state (or ground subspace in a degenerate spectrum) of $H(\tf) = H_1$ in the shortest $\tf$ possible, where $p_{j,t} \coloneqq \bra{E_j^t} \!\rho_t\! \ket{E_j^t}$.

With this setup, we derive a hierarchy of lower bounds on the time $\tf$ needed to perform annealing~\cite{SM},
\begin{align}\label{eq:boundhierarchy}
\tf \geq \tqslone \geq \tqsltwo \geq \tqslthree,
\end{align}
with
\begin{subequations}
\label{eq:QSLtimes}
\begin{align}
\label{eq:QSL1}
\tqslone &\coloneqq 2 \frac{\langle H_0 \rangle_\tf + \langle H_1 \rangle_0 - \langle H_1 \rangle_\tf}{  \big\| [H_1,H_0] \big\| \, \frac{1}{\tf}\int_0^\tf C_1(\rho_t)dt }, \\
\label{eq:QSL2}
\tqsltwo &\coloneqq \frac{\langle H_0 \rangle_\tf + \langle H_1 \rangle_0 - \langle H_1 \rangle_\tf}{  \big\| [H_1,H_0] \big\| \, \frac{1}{\tf} \int_0^\tf \sqrt{1 -  \sum_j p_{j,t}^2} dt },\\
\label{eq:QSL3}
\tqslthree &\coloneqq    \frac{\langle H_0 \rangle_\tf + \langle H_1 \rangle_0 - \langle H_1 \rangle_\tf}{\| [H_1,H_0] \|}.
\end{align}
\end{subequations}
These limits on the
time to reach a solution
through annealing processes constitute the main result of this Letter.

While the first two bounds depend on the trajectory of the state of the system through the annealing process, the loosest of the bounds, $\tqslthree$, only depends on properties of the Hamiltonians $H_0$ and $H_1$, and on how close the final state is to the desired ground state. The error term $\langle H_1 \rangle_\tf$ describes how far the final state is from the desired solution. For perfect annealing, $\langle H_1 \rangle_\tf = 0$.  Note, too, that the second bound implies that an annealing process where the system remains in the instantaneous ground state at all times, $p_{0,t} = 1$, requires an infinite time since $\tqsltwo$ diverges. This is consistent with truly adiabatic evolution.

Alternatively, these bounds set constraints on the minimum coherence and the minimum excitations needed to anneal a system. For concreteness, assume one desires to perfectly anneal a system within a time $\tf$ much shorter than the adiabatic timescale. Equations~\eqref{eq:boundhierarchy}, \eqref{eq:QSL1} and~\eqref{eq:QSL2} then imply that 
\begin{align}
    \int_0^\tf \!\!\!\!\! \sqrt{\vphantom{\sum}\smash{1 - \sum_j p_{j,t}^2}} dt \geq \frac{1}{2}\int_0^\tf \!\!\!\! C_1(\rho_t)dt \geq \frac{\langle H_0 \rangle_\tf + \langle H_1 \rangle_0}{\big\| [H_1,H_0] \big\|}.
\end{align}
While the rightmost term only depends on the Hamiltonians that define the problem, the other two terms are path-dependent. The leftmost term is an entropic quantity that characterizes energy excitations. The middle term depends on the energy coherence of the system. Fast annealing thus requires populating many energy levels. This serves as a sort of converse for the adiabatic theorem, which states that no excitations occur as long as the process is sufficiently slow.
Fast annealing also requires coherence $C_1 > 0$ in the energy basis. This cements the role of coherence as a resource in quantum computations~\cite{CoherenceQCPRA2016,CoherenceQCPRA2017,CoherenceQCPRL2022}.

Next, we prove that these bounds are asymptotically saturable in the size of the system, correctly capturing the optimal annealing timescales of certain toy models.

\emph{Example of optimally fast annealing: unstructured search.---}
Consider the standard model for unstructured search
over $d$ elements
on an analog quantum computer~\cite{RolandPRA2002}. Let the system be initialized in a state $ \ket{\gs(0)} = \ket{\psi_0} = \frac{1}{\sqrt d} \sum_{j=1}^{d} \ket{j}$~\footnote{Interestingly, the starting state is maximally coherent with respect to the target Hamiltonian $H_1$~\cite{CoherencePRL2014}.}, with
\begin{align}\label{eq:unstructured_search}
H_0 = \id - \ket{\psi_0}\!\bra{\psi_0}, \qquad H_1 = \id - \ket{m}\!\bra{m}.
\end{align}
The aim is to find the eigenstate $ |\gs^\tf\rangle \equiv \ket{m}$ among the $d$ possible states. In the limit $d \gg 1$,
Roland and Cerf proved that an optimized adiabatic schedule drives the system to a state that is close to the desired state, with $\big| \langle{\gs^\tf}|  \psi_\tf \rangle \big|^2 \geq 1 - \epsilon^2$, in an adiabatic annealing time $T_{\textnormal{adiab}} = \frac{\pi}{ 2 \epsilon} \sqrt{d}$~\cite{RolandPRA2002}. That is, whereas classically it takes $\sim d$ trials to find an item from an unstructured list, quantum mechanical protocols can do this in a time $\sim \sqrt d$, recovering the $1/\sqrt{d}$ speedup from Grover's algorithm in the digital case.

Using that
\begin{align}
\big\| [H_1,H_0] \big\| = \frac{1}{\sqrt{d}}; \qquad \langle H_1 \rangle_0 = 1 - \frac{1}{d},
\end{align}
and that
\begin{align}
\langle H_0 \rangle_\tf\! - \langle H_1 \rangle_\tf\! = \left| \bra{\psi_\tf\!} m \rangle \right|^2 -  \left| \bra{\psi_\tf\!}  \psi_0 \rangle \right|^2 \geq p_{0,\tf},
\end{align}
we obtain that any annealing protocol requires a time $\tf \nobreak \geq \nobreak \tqslthree$, with
\begin{align}
\tqslthree \geq \frac{1 - \tfrac{1}{d} + p_{0,\tf}}{\tfrac{1}{\sqrt{d}}} \approx 2\sqrt{d}.
\end{align}
That is, the scaling with system size of Roland and Cerf's optimal adiabatic protocol cannot be beaten by diabatic protocols.
This also shows that the lower bound Eq.~\eqref{eq:QSLtimes} on annealing times is (asymptotically) saturable. 

If we further impose, as Roland and Cerf do, that $p_{0,t} \geq 1 - \epsilon^2$ with $\epsilon \ll 1$, we get that
$1 \nobreak-\nobreak \sum_j p_{j,t}^2 \leq \nobreak 1\nobreak -\nobreak p_{0,t}^2 \lesssim \nobreak 2 \epsilon^2$~\footnote{We adopt a notation where $f \gtrsim
a$ ($f \lesssim
a$) means $f > c$ ($f < c$) with $c \sim a$}.
Then, we find that adiabatic annealing requires a time $\tf \nobreak \geq \nobreak \tqsltwo$, where
\begin{align}
\tqsltwo \gtrsim \frac{1 - \tfrac{1}{d} + p_{0,\tf}}{\tfrac{1}{\sqrt{d}} \sqrt{2} \epsilon} \approx \frac{\sqrt{2 d}}{ \epsilon}.
\end{align}
Both the scaling with system size $d$ and target distance $\epsilon$ are saturated by Roland and Cerf's optimal adiabatic protocol, which requires a time $T_{\textnormal{adiab}} = \frac{\pi}{ 2 \epsilon} \sqrt{d}$.

\emph{The gaps between $\tqsl$ and $T_\textnormal{adiab}$.---}
However, as we argued in the introduction, adiabatic schedules can be far from optimal.
In certain models, the gap between the timescales in Eq.~\eqref{eq:QSLtimes} and the ones implied by the adiabatic theorem can be large. In order to explore such gap, we consider two toy models where free parameters  govern the adiabatic timescale $T_\textnormal{adiab}$.  

An example where this is the case is the Hamming spike problem,
defined by the Hamiltonians
\begin{subequations}\label{eq:hammingspike}
\begin{align}
    {H}_0 &= \frac{1}{2}(N\id-M_x), \\ 
   {H}_1 &= \frac{1}{2}(N\id - M_z)+b\left( W\right),
\end{align}
\end{subequations}
where $M_\xi \coloneqq \sum_{\nu = 1}^N \sigma_\nu^\xi$ is the magnetization of the $N$ qubits along direction $\xi \in \{x,y,z \}$, $b$ is a function of the so-called Hamming weight operator $W:=(N\id-M_z)/2$ which, when acting on a computational basis state, returns its Hamming weight $w$, defined as the number of ones in the bit string. We assume $b$ is localized around the region $w=\frac{N}{4}$ and models  a ``spike'' or ``barrier'' of some form~\cite{farhi2002quantum,brady2016spectral, R04, CD14, MAL16, Bringewatt}.  
The barrier is assumed large enough to hinder tunneling of the quantum state during the annealing process, but small enough to act perturbatively. In particular, assume the barrier has height  $\sim N^\alpha $ and width $\sim N^\beta $ with $\alpha<1$ and $\beta<\frac{1}{2}$ ~\cite{brady2016spectral}.
The size of the barrier dictates the timescales derived from the adiabatic theorem. It holds that $T_\textnormal{adiab} \sim \textnormal{poly}(N)$ when $2 \alpha + \beta < 1$, and that $T_\textnormal{adiab} \sim \textnormal{exp}(N)$ for $2 \alpha + \beta > 1$~\cite{brady2016spectral}.

As $\ket{0}$ and $\ket{+}$ are the eigenstates corresponding to the minimum eigenvalues of $-\sigma^z$ and $-\sigma^x$, 
the ground states of $H_0$ and $H_1$ are $\ket{+}^{\otimes N}$ and  $\ket{0}^{\otimes N}$, respectively.
Then, assuming ideal annealing gives
\begin{align}\label{eq:spikeexp}
    \avg{ {H}_0}_\tf &= \frac{N}{2}, \qquad \avg{ {H}_1}_0 = \frac{N}{2}+\mathcal{O}(N^{\alpha+\beta-1/2}),
\end{align}
and it holds that~\cite{SM} 
\begin{align}
\label{eq:spikecomm}
\big\| [ H_0, H_1 ] \big\| &\leq
\frac{N}{2} + \mathcal{O}\!\left(N^{\alpha+\beta}\right),
\end{align}
The important thing to note is that these correction terms depend on the area under the barrier curve $b$ and that in most parameter regimes considered (including some with exponentially small spectral gaps~\cite{brady2016spectral}) they will scale linearly or sub-linearily in $N$. 

Therefore, successful annealing in the Hamming spike problem requires at least a time $\tf \geq \tqslthree$, where
$\tqslthree \gtrsim 1$.
Remarkably, this scaling matches that of the numerically optimized annealing schedules~\cite{LidarHammingSpike,BradyPRA2017}  
and the quantum approximate optimization algorithm (QAOA) schedule for this problems~\cite{BapatHammingSpike}.
This is in stark contrast with the size-dependent timescales obtained from the adiabatic theorem. 
This shows a second toy model where the scaling of the new bounds~\eqref{eq:QSLtimes} is saturated and that the annealing times for the Hamming spike problem previously found numerically in the literature are, in fact, optimal. 

Another toy model with a large gap between adiabatic and non-adiabatic timescales is the $p-$spin model~\cite{PhysRevE.85.051112,Seoane_2012}.
In the ferromagnetic $p-$spin model, 
%\begin{align}\label{eq:pspinmodel}
%H_0 =  \frac{1}{2} (N\id-M_x), \qquad H_1 = \frac{N^{1-p}}{2} ( N^p \id-M_z^p).
%\end{align}
\begin{align}\label{eq:pspinmodel}
H_0 =  \frac{N}{2} \left(\id-\frac{M_x}{N}\right), \qquad H_1 = \frac{N}{2} \left(  \id-\frac{M_z^p}{N^p}\right).
\end{align}
The integer $p \geq 1$ governs the timescales in the adiabatic theorem via the minimum gap which scales as~\cite{bapst2012quantum}
\begin{subequations}
\begin{align}
\Delta \nobreak &\sim \nobreak  1, \qquad p \nobreak = \nobreak 1 \\
\Delta \nobreak &\sim \nobreak  N ^{-1/3}, \qquad p \nobreak = \nobreak 2 \\
\Delta \nobreak &\sim \nobreak   \exp(-N)  ,\qquad p \nobreak \geq \nobreak 3,
\end{align}
\end{subequations}
yielding adiabatic timescales of $T_\textnormal{adiab} \nobreak \sim \nobreak \{ 1, N^{2/3} , \exp(2N) \}$, respectively. 

In contrast, our bound in Eq.~\eqref{eq:QSLtimes} yields
\begin{align}\label{eq:t3pspin}
\tqslthree \geq 2, \qquad  \forall \, p \geq 1,
\end{align}
where we used the fact that the ground states of $H_0$ and $H_1$ are $\ket{+}^{\otimes N}$ and $\ket{0}^{\otimes N}$ (for odd $p$) or $\big\{\ket{0}^{\otimes N},\ket{1}^{\otimes N}\big\}$  (for even $p$)---implying that $\avg{H_0}_\tf = \avg{H_1}_0 = N/2$---and the fact that $ \big\| [H_1,H_0] \big\| \leq N/2$~\cite{SM}.

The outstanding question is: which of these widely different timescales better characterize the performance of an optimal schedule?
For odd $N$, it is known analytically that a constant time, single round QAOA-style, or bang-bang, annealing schedule (with $g_t=1$ for an initial interval of time and $g_t=0$ for the rest) allows one to exactly reach the target ground state~\cite{optimalpferro2020}. Eq.~(\ref{eq:t3pspin}) demonstrates that this scaling is, in fact, optimal. We show a simple proof of this in the supplemental material~\cite{SM}. While analytically less straightforward, numerics for even $N$ also indicate 
$t_f \sim 1$
scaling to reach the target state with high fidelity.

Therefore, we have a third toy model where the optimal schedule saturates the lower bounds $\tqsl$, and where the gap to the adiabatic timescale $T_\textnormal{adiab}$ is large (exponential for $p \geq 3$).
\begin{figure*}
  \centering  
  \includegraphics[trim=00 00 00 00,width=1 \textwidth]{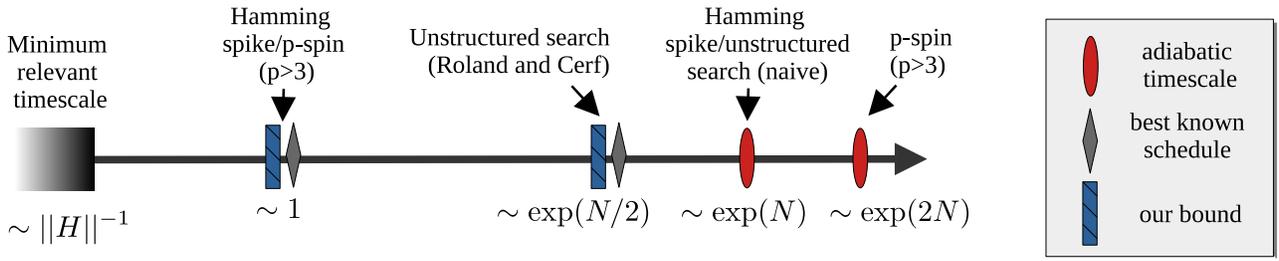} 
\caption{\label{fig:fig1}
 \textbf{Annealing timescales.} An illustration of the range of possible timescales in annealing problems and how our bounds and the adiabatic timescales fit in the picture. 
 }
\end{figure*}

\emph{Lower bounds for $k$-local
Hamiltonians.---}
Consider the $N$-particle Hamiltonians
\begin{align}
H_0 = \sum_{\nu=1}^N h_\nu^0, \qquad H_1 = \sum_{\nu=1}^N h_\nu^1,
\end{align}
where $h_\nu^0$ and $h_\nu^1$ are $k$-local Hamiltonians with support on at most $k$ subsystems~\footnote{We are adopting the naming convention typically used in computer science. Note, though, that these Hamiltonians need not be geometrically local. The role of space locality on the minimum annealing timescales remains an interesting problem to be explored.}, where $\| h_\nu^0 \| \nobreak = \nobreak  \| h_\nu^1 \| \nobreak = \nobreak 1$.  

The scaling with $N$ of the bounds on annealing times, Eq.~\eqref{eq:QSLtimes}, intricately depends on the constituent Hamiltonians. 
However, it holds that $\big\| [ H_0, H_1 ] \big\| \leq k \, N$, and
one can typically expect that $\langle H_0 \rangle_\tf \nobreak \sim \nobreak \langle H_1 \rangle_0 \nobreak \sim \nobreak N $.
This gives that any annealing protocol that aims to connect grounds states of $k$-local Hamiltonians requires a time $\tf \geq \tqslthree$ with
\begin{align}
\label{eq:boundManyBody}
\tqslthree \gtrsim \frac{2}{k}.
\end{align}

This scaling is in stark contrast with the one obtained from the adiabatic theorem. For many-body systems, the minimum energy gap between the ground and first excited state typically scale as $\Delta \nobreak \sim \nobreak 1/\textnormal{poly}(N)$ or as $\Delta \nobreak \sim \nobreak \exp({-N})$~\cite{GapsShor}. In the latter case, for example, the adiabatic theorem ensures a schedule that anneals the system if $\tf > T_\textnormal{adiab}$ with 
\begin{align}
T_\textnormal{adiab} \sim  \frac{\theta}{\Delta^2}  \sim \exp(2N).
\end{align}

This gives the same scaling as, for instance, the $p$-ferromagnetic spin model for $p = 3$, which is $3$-local. However, in that case we found that the scaling of lower bound $\tqslthree \geq 2$ was saturated by a single-round QAOA schedule. This thus shows that the minimum annealing time Eq.~\eqref{eq:boundManyBody} for $k$-local systems is indeed saturable.

\emph{Annealing times with extra control Hamiltonians.---}
So far, we adopted the standard quantum annealing scenario where one carefully tailors a schedule that combines $H_0$ and $H_1$ to reach the desired state.
However, including extra control Hamiltonians $H_C$ adds freedom to the dynamics that can, in principle, speedup an annealing process~\cite{Hormozi2017,Susa2017,Vinci2017,Crosson2020,GeneticAlgo}. 
One extreme example of this is that of a counterdiabatic Hamiltonian $H_{CD}$ that implements shortcut to adiabaticity dynamics by inhibiting excitations out of the instantaneous ground state~\cite{berry2009transitionless,
AdCshortcutsPRL2013,
ShortcutsMugaRevModPhys2019,
AdC2019focus,
PhysRevResearch.2.013283}.
%(see applications to quantum annealing in Refs.~\cite{ShortcutsAnnealingPRA2017,ShortcutsAnnealingPRApplied2021,ShortcutsAnnealingPRA2022}).

How much can extra physical control Hamiltonians speedup an annealing process? Let us assume access to a set of $N_C$ control Hamiltonians $\{ H_C^a \}$ with schedules $\{ f_t^a \} \geq 0$ such that $f_0^a = f_\tf^a = 0$. Their aim is to speedup the transition to the eigenstate $\ket{\gs(\tf)}$ of $H_1$. The total Hamiltonian is
\begin{align}
\widetilde{H}(t) = H(t) + \sum_{a=1}^{N_C} f_t^a H_C^a.
\end{align}
Then, we prove
a constraint on the annealing times $\tf \nobreak \geq \nobreak \tqsl$ under dynamics with the extra control knobs provided by $\{ H_C^a \}$, where~\cite{SM}
\begin{align}
\label{eq:QSLControl}
\tqsl \coloneqq \frac{ \langle H_0 \rangle_\tf + \langle H_1 \rangle_0 - \langle H_1 \rangle_\tf }{\Big\| \big[ H_1,H_0\big] \Big\| + \sum_{a=1}^{N_C} \Big\| \big[ H_1-H_0, H_C^a\big] \Big\|}.
\end{align}
Consequently, while a control Hamiltonian $H_C \nobreak \propto \nobreak  (H_1 \nobreak - \nobreak H_0)^p$ may improve the performance of some schedules, it cannot improve the performance of an optimal schedule that saturates the lower bounds~\eqref{eq:QSLtimes}. Some interesting connections can also be made between Eq.~(\ref{eq:QSLControl}) and counterdiabatic Hamiltonians.
Note that $H_1-H_0=\dot H(t)/\dot g$. Using Eqs.~(3, 4) in Ref.~\cite{Petiziol2018}, this leads to $[H_1-H_0, H_{CD}] = -(1/\dot g) \sum_{j \neq k} A_{jk} \ket{E_j^t}\!\bra{E_k^t}$, where $A_{jk} \geq 0$. This in turn typically leads to $[H_1-H_0, H_{CD}]$ with a large norm. Consequently, if one catalyzes the evolution with an additional control Hamiltonian that is equivalent to the counterdiabatic Hamiltonian, this will lead to a sizeable increase in the denominator of Eq.~(\ref{eq:QSLControl}) and, therefore, a reduction in the lower bound on the annealing time.

 Finally, Eq.~\eqref{eq:QSLControl} implies a lower limit on the number of control Hamiltonians $N_C$ needed to perform annealing within a short time $\tf$. For instance, assume that one desires to anneal the system in the maximum timescale $1/\|H(t)\|$ possible given the original Hamiltonian $H(t)$ in Eq.~\eqref{eq:Ham}. For concreteness, let us assume, as in the $k-$local Hamiltonians above, that $\langle H_0 \rangle_\tf \sim \langle H_1 \rangle_0 \sim N$ and that the control Hamiltonians are also $k-$local, so that $\big\| [ H_1 - H_0 , H_C^a ] \big\| \leq k N$. Then, 
\begin{align}
\| H(t) \| \, \tqsl \geq N \frac{2N}{kN (1 + N_C)} \sim \frac{N}{k N_C},
\end{align}
and at least $N_C \sim N/k$ control Hamiltonians are needed to perform annealing at the maximum rate $\| H(t) \|$ defined by the original Hamiltonian.
Similarly, 
Eq.~\eqref{eq:QSLControl} implies that at least $N_C \sim \frac{N}{k \tf}$ control Hamiltonians are needed to implement a counterdiabatic Hamiltonian that enforces adiabatic evolution in $\tf$ for a many-body system.

\emph{Discussion.---} Extensive work has been devoted to understanding the timescales $T_\textnormal{adiab}$ that ensure that a process is adiabatic.
However, less is rigorously known about diabatic schedules that can anneal a system faster.  In fact, to our knowledge the scaling of the optimal annealing time is only known in few toy models, which include the unstructured search model, the Hamming spike problem, and the $p-$ferromagnetic spin model.

In this Letter, we derived easy-to-evaluate lower bounds on the times necessary for annealing to occur which are saturated by the best known annealing schedules for all of these toy models. While the Roland and Cerf schedule appears to have optimal scaling even without confirmation from our bounds due to the fact it recovers the Grover-type speedup for unstructured search~\footnote{Note that this argument is not rigorous, as it has been argued that analog classical models can also provide this same speedup, calling into question whether the speedup in the Roland and Cerf model is genuinely due to quantum effects or just due to finely tuned analog control~\cite{hen2019quantum, slutskii2019analog}. Either way, our results show that the schedule achieves optimal scaling given access to $H_0$ and $H_1$.}, studies of the Hamming spike problem were numerical in nature. Moreover, we found that previously considered QAOA schedules for the $p-$ferromagnetic spin model also saturate the lower bounds, proving those schedules to be optimal.

Note that all models considered here are Hamming symmetric.
That is, the Hamiltonians are invariant under permutations of basis elements with the same Hamming weight in the computational basis, or equivalently, they conserve the total spin along a given ($z$) direction.
This high degree of symmetry could conceivably be responsible for the saturation of our bounds in these models. Strikingly, however, the collection of models for which we can show our bounds are saturable exhibit vastly different optimal annealing schedules, ranging from optimized adiabatic schedules to ``bang-bang'' controls. This means that, if symmetry is indeed responsible for the tightness of the bounds, the direct means by which it causes this tightness is not obvious. We leave exploring this as an open question, while observing that the variety of different schedules which saturate bounds provide compelling evidence for their usefulness, especially given the dearth of rigorous results on
 the timescales needed to perform 
quantum annealing beyond the adiabatic regime. 

Unlike the timescales obtained from the adiabatic theorem, our bounds do not depend on the spectral gap of the system, which makes it easier to evaluate the latter. While we found our bounds to better reflect the timescales of optimized annealing in all toy models considered, this highlights the importance of understanding the role of the spectral gap in the performance of optimal schedules in more physically realistic scenarios.
In addition, because our bounds involve the quantum coherence of the system, this suggests an approach to understanding when the system can escape from local minima~\cite{altshuler2010anderson} in the diabatic regime.
Finally, the role that geometric locality plays in the lower bounds on annealing times remains a problem to be explored.

\noindent \emph{Acknowledgements.---}
We thank Michael Gullans for discussions related to this work.
We acknowledge funding by the U.S. Department of Energy (DOE) ASCR Accelerated Research in Quantum Computing program (award No.~DE-SC0020312), DoE QSA, NSF QLCI (award No.~OMA-2120757), DoE ASCR Quantum Testbed Pathfinder program (award No.~DE-SC0019040), NSF PFCQC program, AFOSR, ARO MURI, AFOSR MURI, and DARPA SAVaNT ADVENT.
This research was supported in part by the National Science Foundation under Grant No. NSF PHY-1748958, the Heising-Simons Foundation, and the Simons Foundation (216179, LB).
The work at Los Alamos National Laboratory was carried out under the auspices of the US DOE and NNSA under contract No.~DEAC52-06NA25396.
We also acknowledge support by the DOE
Office of Science, Office of Advanced Scientific Computing Research, Accelerated Research for Quantum Computing program, Fundamental Algorithmic Research for Quantum Computing (FAR-QC) project.
L.~T.~B. is a KBR employee working under the Prime Contract No. 80ARC020D0010 with the NASA Ames Research Center and is grateful for support from the DARPA RQMLS program under IAA 8839, Annex 128.
The United States Government retains, and by accepting the article for publication, the publisher acknowledges that the United States Government retains, a nonexclusive, paid-up, irrevocable, worldwide license to publish or reproduce the published form of this work, or allow others to do so, for United States Government purposes.

\bibliography{referencesannealing}

\clearpage
\newpage
\onecolumngrid

\section*{Supplemental Material}
\setcounter{secnumdepth}{1}
\renewcommand{\thesection}{S\arabic{section}}
\setcounter{equation}{0}
\renewcommand{\theequation}{S\arabic{equation}}

In this supplemental material, we provide details for a number of statements in the main text. In Section~\ref{s:lower_bounds}, we give the full derivation for our lower bounds on annealing times. In Sections~\ref{s:hamming} and \ref{s:pspin} we provide algebraic details for the Hamming spike example and $p$-spin model, respectively. In Section~\ref{s:control_hamiltonians}, we provide derivations for results related to having extra control Hamiltonians.

\section{Lower bounds on annealing times}\label{s:lower_bounds}

Here, we derive the lower bounds on annealing times. Using that $\langle H_0 \rangle_t \coloneqq \tr{\rho_t H_0}$ and $\langle H_1 \rangle_t \coloneqq \tr{\rho_t H_1}$, it holds that
\begin{align}
\label{eq-app:initial_eq}
i \tr{\rho_t [H_1,H_0]} = \frac{d \langle H_0 \rangle_t}{dt} - \frac{d \langle H_1 \rangle_t}{dt}.
\end{align}

Integrating over the duration of the annealing schedule gives 
\begin{align}
i \int_0^\tf \tr{\rho_t [H_1,H_0]} dt &= \langle H_0 \rangle_\tf + \langle H_1 \rangle_0 - \langle H_1 \rangle_\tf,
\end{align}
where without loss of generality we assume that the ground states of $H_0$ and $H_1$ are zero. 
Then, 
\begin{align}
\label{eq-app:exacttime}
\tf = \frac{\langle H_0 \rangle_\tf + \langle H_1 \rangle_0 - \langle H_1 \rangle_\tf}{i \tr{\overline \rho [H_1,H_0]}},
\end{align}
where $\overline \rho \coloneqq \frac{1}{\tf} \int_0^\tf \rho_t dt$ is the time averaged state of the system.

Since Eq.~\eqref{eq-app:exacttime} is exact, solving for $\tf$ is as hard as solving the dynamics of the system. 
However, we can obtain lower bounds on $\tf$ by upper bounding the denominator. Note that
\begin{align}
\label{eq-app:auxaveragecommut}
i \tr{\overline \rho [H_1,H_0]}  &= \frac{i}{\tf} \int_0^\tf \tr{\rho_t \frac{1}{g_t}[(1-g_t)H_0 + g_t H_1,H_0]} dt   \nonumber \\
&= \frac{i}{\tf} \int_0^\tf \tr{\frac{1}{g_t}[\rho_t  ,H_t] H_0} \frac{dt}{g_t}   \nonumber \\
&= \frac{i}{\tf} \int_0^\tf \tr{\frac{1}{g_t}[\rho_t - \sigma_t  ,H_t] H_0} \frac{dt}{g_t} \nonumber \\
&= \frac{i}{\tf} \int_0^\tf \tr{\left(\rho_t - \sigma_t  \right)  [H_1,H_0]} dt \nonumber \\
&= \frac{i}{\tf} \int_0^\tf \min_{\sigma_t} \tr{\left(\rho_t - \sigma_t  \right)  [H_1,H_0]} dt.
\end{align}
Here, $\sigma_t$ is a state that commutes with the Hamiltonian $H_t = \sum_j E_j^t \ket{E_j^t}\!\bra{E_j^t}$.

For any Hermitian operator $A$, it holds that
\begin{subequations}
\label{eq-app:auxineq}
\begin{align}
\min_{\sigma_t} \, \tr{ \left( \rho_t -  \sigma_t \right)  A }  
&\leq  \frac{1}{2} \| A \| \, \min_{\sigma_t} \, \left\| \rho_t -  \sigma_t \right\|_1 = \frac{1}{2} \| A \| \,  C_1(\rho_t) \\
&\leq  \frac{1}{2} \| A \| \, \left\| \rho_t -  \mathcal{D}_t(
\rho_t) \right\|_1 \leq \| A \| \,  \sqrt{1 - \sum_j p_{j,t}^2} \\
&\leq  \| A \|.
\end{align}
\end{subequations} 
In the first line, we use that $\tr{A (\rho-\sigma)} \nobreak \leq \nobreak \tfrac{1}{2} \| \rho \nobreak - \nobreak \sigma \|_1 \, \|A\| $, where $\| A \|_1 \nobreak \coloneqq \nobreak \tr{ \sqrt{A A^\dag} }$ is the trace norm and $\| A \|$ is the largest singular value of an operator $A$~\cite{nielsen_chuang_2010} and that $C_1(\rho_t) \coloneqq \min_{\sigma_t} \big\| \rho_t - \sigma_t \big\|_1$.
For the second line, we use that $C_1(\rho_t) \leq \big\| \rho_t - \mathcal{D}_t(\rho_t) \big\|_1$ where $\mathcal{D}_t(\rho_t) \coloneqq \sum_j p_{j,t}\ket{E_j^t}\!\bra{E_j^t}$ is the state dephased in the energy basis and that
$\| \rho_t-\mathcal{D}_t(\rho_t) \|_1 \nobreak \leq \nobreak 2 \sqrt{1 - \tr{\rho_t \mathcal{D}_t(\rho_t)}}$ given that $\rho_t$ is pure~\cite{wilde_2017}.
The last inequality follows from $p_{j,t} \geq 0$ for all $j,t$.

Combining Eqs.~\eqref{eq-app:exacttime} and~\eqref{eq-app:auxaveragecommut} with the last three inequalities in Eq.~\eqref{eq-app:auxineq} yields a hierarchy of lower bounds on the time $\tf$ needed to perform annealing,
\begin{align}
\tf \geq \tqslone \geq \tqsltwo \geq \tqslthree,
\end{align}
with
\begin{subequations}
\label{eq-app:QSLtimes}
\begin{align}
\label{eq-app:QSL1}
\tqslone &\coloneqq 2 \frac{\langle H_0 \rangle_\tf + \langle H_1 \rangle_0 - \langle H_1 \rangle_\tf}{  \big\| [H_1,H_0] \big\| \, \frac{1}{\tf}\int_0^\tf C_1(\rho_t)dt } \\
\label{eq-app:QSL2}
\tqsltwo &\coloneqq \frac{\langle H_0 \rangle_\tf + \langle H_1 \rangle_0 - \langle H_1 \rangle_\tf}{  \big\| [H_1,H_0] \big\| \, \frac{1}{\tf} \int_0^\tf \sqrt{1 -  \sum_j p_{j,t}^2} dt }\\
\label{eq-app:QSL3}
\tqslthree &\coloneqq    \frac{\langle H_0 \rangle_\tf + \langle H_1 \rangle_0 - \langle H_1 \rangle_\tf}{\| [H_1,H_0] \|}.
\end{align}
\end{subequations}

\section{Hamming spike problem}\label{s:hamming}
In this section, we consider the Hamming spike problem defined in Eq.~(\ref{eq:hammingspike}) of the main text. Let $W:=\sum_{\nu=1}^N (\id \nobreak - \nobreak \sigma^z_\nu)/2$ be the Hamming weight operator. Then, making use of magnetization operators defined in the main text we have (under the assumption of ideal annealing) that
\begin{align}
    \avg{ {H}_0}_\tf&=\bra{0}^{\otimes N}\left(\frac{1}{2}\sum_{\nu=1}^N (\id-\sigma_\nu^x )\right) \ket{0}^{\otimes N} \nonumber \\
    &=\frac{1}{2}\bra{0}^{\otimes N}\left(N\id - M_x\right)\ket{0}^{\otimes N} \nonumber\\
    &=\frac{N}{2},
\end{align}
and that
\begin{align}
    \avg{ {H}_1}_0&=\bra{+}^{\otimes N}\left(\frac{1}{2}\sum_{\nu=1}^N (\id-\sigma_\nu^z )+b(W)\right) \ket{+}^{\otimes N} \nonumber\\
    &=\frac{1}{2}\bra{+}^{\otimes N}\left(N\id - M_z+2b(W)\right)\ket{+}^{\otimes N}  \nonumber\\
    &=\frac{N}{2}+\bra{+}^{\otimes N}b(W)\ket{+}^{\otimes N} \nonumber\\
    &=\frac{N}{2}+\mathcal{O}(N^{\alpha+\beta-1/2}).
\end{align}
The correction factor in the last line comes from the fact that $b(W)$ is diagonal with $\sim N^\beta\binom{N}{N/4}\sim 2^N N^{\beta-1/2}$ non-zero elements of magnitude $\sim N^\alpha$ and $\ket{+}^{\otimes N}$ is an even superposition over bit strings with probability amplitudes $2^{-N/2}$.  Given our assumptions that $\alpha<\frac{1}{2}$ and $\beta<1$, this correction factor is at most linear in $N$, so
\begin{equation}
    \avg{ {H}_1}_0\sim N
\end{equation}

Making use of the fact that $[M_x,M_z] = -2i M_y$, we also have that
\begin{align}
    \big\| [H_1,H_0] \big\| &= \frac{1}{4} \big\|  [M_z,M_x]+ 2[b(W), M_x]  \big\|  \nonumber\\
    &\leq \frac{1}{4} \big\|  [M_z,M_x]\big\|+ \frac{1}{2}\big\| [b(W), M_x]  \big\| \label{eq:triangle_ineq_1} \\
    &=  \frac{1}{2}\left( \big\| M_y\big\|+ \big\| [b(W), M_x]  \big\| \right) \nonumber\\
    &\leq \frac{N}{2} + \frac{1}{2}\big\| [b(W), M_x]  \big\|,
\end{align}
where we used the triangle inequality in Eq.~(\ref{eq:triangle_ineq_1}). We can further bound the second term by noting that $b(W)$ is diagonal in the computational basis $\{\ket{0},\ket{1}\}$. Therefore,
\begin{align}
    [b(W), M_x]_{ij}=\left((b(W)_{ii}-b(W)_{jj}\right)(M_x)_{ij}=\begin{cases}
    0, & \text{if } w_i=w_j\\
    \sim N^\alpha, & \text{otherwise}.
    \end{cases}
\end{align}
Here, we let $w_i, w_j$ denote the Hamming weight of the computational basis states indexed by $i,j$, respectively and used the Hamming symmetry of $b(W)$. We know that the width (in Hamming weight) of the barrier specified by $b(W)$ is $\sim N^\beta$ so there are $\sim N^\beta$ non-zero matrix elements in the commutator $[b(W), M_x]$. Consequently,
\begin{align}
    \big\| [H_1,H_0] \big\| &\leq \frac{N}{2} + \mathcal{O}\left(N^{\alpha+\beta}\right).
\end{align}

\section{Ferromagnetic p-spin model}\label{s:pspin}

\subsection{Bound on the commutator norm}
Using that $[AB,C]=A[B,C]+[A,C]B$ and $[M_x,M_z] = -2i M_y$, we have that for the ferromagnetic p-spin model given in Eq.~(\ref{eq:pspinmodel}):
\begin{align}
\big\| [H_1,H_0] \big\| &=  \frac{N^{1-p}}{4} \big\|  [M_z^p,M_x]   \big\|  \nonumber\\
&= \frac{N^{1-p}}{4} \Big\| M_z^{p-1}[M_z,M_x] + [M_z^{p-1},M_x]M_z \Big\|  \nonumber\\
&= \frac{N^{1-p}}{4} \Big\| M_z^{p-1}[M_z,M_x] + M_z^{p-2}[M_z,M_x]M_z + [M_z^{p-2},M_x]M_z^2  \Big\|  \nonumber\\
&= \frac{N^{1-p}}{2} \Big\| M_z^{p-1}M_y+ M_z^{p-2}M_yM_z + M_z^{p-3}M_yM_z^2+\cdots+M_yM_z^{p-1} \Big\|  \nonumber\\
&\leq \frac{N^{1-p}}{2} N^p \label{eq:triangle_ineq}\\
&=\frac{N}{2},
\end{align}
where in Eq.~(\ref{eq:triangle_ineq}) we used the triangle inequality and that a string of $p$ magnetization operators has a norm bounded by $N^p$.

\subsection{Optimal annealing schedule for odd $N$}
In the main text, we stated the result from Ref.~\cite{optimalpferro2020} that the ferromagnetic $p$-spin model for odd $N$ allows for a constant time annealing schedule obtaining the exact target ground state. Here, for completeness, we provide a short explanation of our own for this result, highlighting a geometric interpretation of the behavior of the algorithm.

For simplicity of this presentation we also restrict ourselves to $p$ odd. Here, the optimal schedule is particularly simple: a one round QAOA-style protocol with $g_t=1$ for $t\in[0,\frac{\pi}{4})$ and $g_t=0$ for $t\in[\frac{\pi}{4},\pi]$. Observing that the initial state in the $p$-spin model is $\ket{+}^{\otimes n}$ and the target state is $\ket{0\cdots 0}$ so this schedule is simply the following (up to irrelevant global phases):
\begin{equation}
    \ket{0\cdots 0}=e^{i\frac{\pi}{4}M_x}e^{i\frac{\pi}{4}M_z^p}\ket{+}^{\otimes n}
\end{equation}

The reason this protocol works is that for odd $p$
\begin{equation}\label{eq:pi4sc}
    e^{i\frac{\pi}{4}M_z^p}=e^{i\frac{\pi}{4}M_z}.
\end{equation}
Therefore, geometrically, the protocol can be viewed in terms of $SU(2)$ spin-coherent states as a rotation from a spin-coherent state along the $M_x$ axis to a state along the $M_y$ axis, followed by a rotation up to the $M_z$ axis. Importantly, however, the dynamics are not confined to this two-dimensional subspace of spin coherent states, as  $e^{i\theta M_z^p}\neq e^{i\theta M_z}$ for arbitrary $\theta$. It is a unique feature of $p$ odd and the choice of $\theta=\pi/4$ that causes the state to return to the space of spin coherent states. We now show this. 

Consider expanding the left hand side of Eq.~(\ref{eq:pi4sc}). We can use that $M_z$
 is diagonal in the computational basis state to restrict our attention to the diagonal terms. Then,
 \begin{equation}
     \left(e^{i\frac{\pi}{4}M_z^p}\right)_{aa}=\cos\left(\frac{\pi}{4}w_a^p\right)+i\sin\left(\frac{\pi}{4}w_a^p\right),
 \end{equation}
 where $a\in\{0,1\}^N$ labels a computational basis state and $w_a=N-2|a|$, where $|a|$ is the Hamming weight of $a$. Consequently, for odd $N$, $w_a$ is odd for all $a$. As $w_a$ is odd we can write
 \begin{align}
     w_a^p&=w_aw_a^{p-1} \nonumber\\
     &=w_a\left(w_a^{\frac{p-1}{2}}\right)^2 \nonumber\\
     &=w_a(2k+1)^2 \nonumber\\
     &=w_a\left[4k(k+1)+1\right]
 \end{align}
 for $k\in\mathbb{Z}$, where we used that $\frac{p-1}{2}\in\mathbb{Z}$ and an odd number raised to any power is odd. Consequently, 
 \begin{equation}
     \frac{w_a^p}{4}=k(k+1)w_a+\frac{w_a}{4}.
 \end{equation}
 The term $k(k+1)w_a$ is even as the product of two odd numbers and an even number is even. Therefore,
 \begin{align}
     \left(e^{i\frac{\pi}{4}M_z^p}\right)_{aa}&=\cos\left(\pi\left[k(k+1)w_a+\frac{w_a}{4}\right]\right)+i\sin\left(\pi\left[k(k+1)w_a+\frac{w_a}{4}\right]\right) \nonumber\\
     &=\cos\left(\frac{\pi}{4}w_a\right)+i\sin\left(\frac{\pi}{4}w_a\right)= \left(e^{i\frac{\pi}{4}M_z}\right)_{aa},
 \end{align}
 for all $a$, demonstrating Eq.~(\ref{eq:pi4sc}). For even $p$, a similar approach works, but the angle of the first rotation is no longer $\pi/4$ and is $N$-dependent. We refer the reader to Ref.~\cite{optimalpferro2020} for details.

\section{Lower bounds on annealing times with extra control Hamiltonians}\label{s:control_hamiltonians}

%\subsection{Lower bounds on annealing times}

When evolving under \begin{align}
\widetilde{H}(t) = H(t) + \sum_{a=1}^{N_C} f_t^a H_C^a,
\end{align}
it holds that
\begin{align}
\frac{d \langle H_0 \rangle_t}{dt} - \frac{d \langle H_1 \rangle_t}{dt}
&=   -i \tr{[H(t),\rho_t] (H_0 - H_1)} - i \sum_{a=1}^{N_C} f_t^a \tr{ [H_C^a,\rho_t]     (H_0 - H_1) }  \nonumber \\
&= i\tr{\rho_t [H_1,H_0]}  + i \sum_{a=1}^{N_C} f_t^a \tr{\rho_t [H_1-H_0,H_C^a]}.
\end{align}
Averaging over the duration $\tf$ of the schedule gives
\begin{align}
   \frac{\langle H_0 \rangle_\tf + \langle H_1 \rangle_0 - \langle H_1 \rangle_\tf}{\tf} =  i\tr{\overline{\rho} [H_1,H_0]} +  i \sum_{a=1}^{N_C} \frac{1}{\tf}\int_0^\tf f_t^a \tr{\rho_t [H_1-H_0,H_C^a]}.
\end{align}
Using that $\left| \tr{\sigma A} \right| \leq \|\sigma\|_1 \|A\|$, $\| \overline{\rho} \|_1 = 1$, the triangle inequality, and that $f_t^a \geq 0$, we get
\begin{align}
   \frac{\langle H_0 \rangle_\tf + \langle H_1 \rangle_0 - \langle H_1 \rangle_\tf}{\tf} &\leq \left\| [H_1,H_0] \right\| + \sum_{a=1}^{N_C} \left\| \tfrac{1}{\tf} \int_0^\tf f_t^a \rho_t \right\|_1 \, \left\|  [H_1 - H_0, H_C^a]  \right\| \nonumber \\
   &\leq \left\| [H_1,H_0] \right\| + \sum_{a=1}^{N_C} \tfrac{1}{\tf} \int_0^\tf f_t^a \left\| \rho_t \right\|_1 \, \left\|  [H_1 - H_0, H_C^a]  \right\| \nonumber \\
    &\leq \left\| [H_1,H_0] \right\| + \sum_{a=1}^{N_C}  \left\|  [H_1 - H_0, H_C^a]  \right\|.
\end{align}
Thus, 
\begin{align}
    \tf \geq \tqsl \coloneqq \frac{\langle H_0 \rangle_\tf + \langle H_1 \rangle_0 - \langle H_1 \rangle_\tf}{  \left\| [H_1,H_0] \right\| + \sum_{a=1}^{N_C} \left\|  [H_1 - H_0, H_C^a]  \right\|  },
\end{align}
which proves Eq.~\eqref{eq:QSLControl} in the main text.

\end{document}